\documentclass[%
reprint,
superscriptaddress,
nofootinbib,
amsmath,
amssymb,
aps,
linenumbers,
prl,
twocolumn
]{revtex4}

\usepackage{graphicx}
\usepackage{dcolumn}
\usepackage{bm}
\usepackage{hyperref}
\usepackage{xcolor}
\usepackage{subcaption}
\usepackage{orcidlink}
\usepackage{txfonts}
\usepackage{natbib}
\usepackage{booktabs}
\usepackage{mathtools}
\usepackage{ragged2e}

\usepackage{caption}

\begin{document}

\preprint{APS/123-QED}

\title{Measurement of the Cosmic Ray  Nickel Energy Spectrum from 10 GeV/n to 2 TeV/n with the DAMPE Space  Mission}

\clearpage

\author{F.~Alemanno,\orcidlink{0000-0003-2021-9205}}
\affiliation{Dipartimento di Matematica e Fisica E. De Giorgi, Universit\`a del Salento, I-73100, Lecce, Italy}
\affiliation{Istituto Nazionale di Fisica Nucleare (INFN) - Sezione di Lecce, I-73100, Lecce, Italy}

\author{Q.~An}
\altaffiliation[]{Deceased}
\affiliation{State Key Laboratory of Particle Detection and Electronics, University of Science and Technology of China, Hefei 230026, China} 
\affiliation{Department of Modern Physics, University of Science and Technology of China, Hefei 230026, China}
\author{P.~Azzarello}
\affiliation{Department of Nuclear and Particle Physics, University of Geneva, CH-1211, Switzerland}

\author{F.~C.~T.~Barbato,\orcidlink{0000-0003-0751-6731}}
\affiliation{Gran Sasso Science Institute (GSSI), Via Iacobucci 2, I-67100 L'Aquila, Italy}
\affiliation{Istituto Nazionale di Fisica Nucleare (INFN) - Laboratori Nazionali del Gran Sasso, I-67100 Assergi, L'Aquila, Italy}

\author{P.~Bernardini,\orcidlink{0000-0002-6530-3227}}
\affiliation{Dipartimento di Matematica e Fisica E. De Giorgi, Universit\`a del Salento, I-73100, Lecce, Italy}
\affiliation{Istituto Nazionale di Fisica Nucleare (INFN) - Sezione di Lecce, I-73100, Lecce, Italy}

\author{X.~J.~Bi}
\affiliation{University of Chinese Academy of Sciences, Beijing 100049, China}
\affiliation{Particle Astrophysics Division, Institute of High Energy Physics, Chinese Academy of Sciences, Beijing 100049, China}

\author{H.~V.~Boutin,\orcidlink{0009-0004-6010-9486}}
\affiliation{Department of Nuclear and Particle Physics, University of Geneva, CH-1211, Switzerland}

\author{I.~Cagnoli,\orcidlink{0000-0001-8822-5914}}
\affiliation{Gran Sasso Science Institute (GSSI), Via Iacobucci 2, I-67100 L'Aquila, Italy}
\affiliation{Istituto Nazionale di Fisica Nucleare (INFN) - Laboratori Nazionali del Gran Sasso, I-67100 Assergi, L'Aquila, Italy}

\author{M.~S.~Cai,\orcidlink{0000-0002-9940-3146}}
\affiliation{Key Laboratory of Dark Matter and Space Astronomy, Purple Mountain Observatory, Chinese Academy of Sciences, Nanjing 210023, China}
\affiliation{School of Astronomy and Space Science, University of Science and Technology of China, Hefei 230026, China}

\author{E.~Casilli,\orcidlink{0009-0003-6044-3428}}
\affiliation{Gran Sasso Science Institute (GSSI), Via Iacobucci 2, I-67100 L'Aquila, Italy}
\affiliation{Istituto Nazionale di Fisica Nucleare (INFN) - Laboratori Nazionali del Gran Sasso, I-67100 Assergi, L'Aquila, Italy}

% ...

\collaboration{DAMPE Collaboration}
\altaffiliation{dampe@pmo.ac.cn}
\date{\today}

\begin{abstract}
Nickel, one of the most tightly bound nuclei alongside iron, is the most abundant heavy element beyond iron in cosmic rays. With DAMPE’s excellent charge resolution and broad energy range, a high-precision energy spectrum provides valuable insights into the acceleration sources of heavy nuclei and their propagation through the interstellar medium. In this analysis, we report the direct measurement of cosmic-ray nickel spectrum from 10 GeV/n to 2 TeV/n with nine years of flight data. The nickel spectrum is consistent with a single power law with spectral index $-2.60 \pm 0.03$ from 40 GeV/n to 1 TeV/n. This work provides an accurate measurement of differential flux of nickel with kinetic energy extending to TeV/n for the first time. 

\end{abstract}
\maketitle
\normalsize

\emph{Introduction}-Galactic cosmic rays, discovered by Hess in 1913,  are high-energy particles originating from outer space and traversing the Galaxy. They can serve as valuable probes for exploring astrophysical particle accelerators and the interstellar medium of the Galaxy~\cite{2015Grenier}. Traditionally, the energy spectra of cosmic-ray nuclei has been assumed to be a single power law~\cite{1949Fermi}. However, with the accumulation of observed data and the expansion of the energy range, many experiments indicate that the spectra of cosmic rays exhibit deviations from a single power law in the energy range around hundreds of GeV/n for proton, helium, and other heavy primary nuclei in cosmic rays~\cite{2009Panov,2017Yoon,2011Adriani,2013Adriani,2015Aguilar,2020Aguilar,2015Aguilar_1,2017Aguilar,2021Aguilar,DAMPEProton,DAMPEHe,CALETProton,CALETIron}.

Heavy nuclei observed in cosmic rays primarily originate from stellar nuclear fusion processes~\cite{Synthesis1}. Iron, with the highest binding energy per nucleon among all stable nuclei, is the most abundant element among nuclei heavier than silicon. The spectrum shape around TeV/n of iron was reported by several experiments such as AMS-02~\cite{2021Aguilar}, CALET~\cite{CALETIron}, and NUCLEON~\cite{NUCLEONIron}. Although there are several theoretical models about the synthesis, acceleration, and propagation of heavy nuclei in cosmic rays~\cite{Synthesis1,Synthesis2,Synthesis3,Spectral1,Spectral2}, current observational results for nickel and heavier nuclei mainly focus on relative abundance measurements~\cite{Supertiger,Cris}.  Even about nickel, the most abundant species in this trans-iron group, only a few experiments, such as CRISIS~\cite{1981Cris}, HEAO3-C2~\cite{1990HEAO3} and G. Minagawa~\cite{1981Balloon}, provided differential energy spectrum in the low energy region below 40 GeV/n. Besides these, the CALET experiment reported the energy spectrum from 8.8 GeV/n to 240 GeV/n~\cite{CALETNickel}, while the NUCLEON reported the upper limit at 511 GeV/n~\cite{NUCLEON}. Consequently, a new precision measurement may clarify the behavior of nickel flux at the highest energies.   

The Dark Matter Particle Explorer (DAMPE, also known as "Wukong" in China), is a calorimetric-type, satellite-borne detector with the goal to observe the high-energy electrons, gamma rays, and cosmic rays~\cite{dampemission1}, launched into a 500-km Sun-synchronous orbit on December 17, 2015. From top to bottom, DAMPE consists of a plastic scintillator detector (PSD)~\cite{psd,orbitdata4,2019Tiekuang,psd2}, a sillicon-tungsten tracker converter (STK)~\cite{stk1,stk2}, a bismuth germanate (BGO) imaging calorimeter~\cite{bgo1,bgo2}, and a neutron Detector (NUD)~\cite{nud}. The PSD measures the charges of incident particles and contributes to the anticoincidence measurement of gamma rays. The STK reconstructs the trajectory and provides additional charge measurements for light nuclei. The BGO calorimeter measures the energy and trajectory of incident particles and discriminates electrons and photons effectively from hadrons based on their shower profiles. The NUD is used for additional electron and proton discrimination. The calibration with nine-year on-orbit data reveals that DAMPE has excellent stability~\cite{orbitdata1,orbitdata2,orbitdata3,orbitdata5}. Based on this excellent performance, combined with its large geometric factor and wide energy coverage, DAMPE is capable of extending measurements of cosmic ray nickel and heavier nuclei well into the TeV/n energy range. 

In this Letter, the measurement of the nickel spectrum with kinetic energy from 10 GeV/n to 2 TeV/n is reported. Our results are the first to extend the energy coverage for the nickel spectrum up to TeV/n.

\emph{Data analysis}-This analysis utilized nine years of data collected by DAMPE between January 1, 2016 and December 31, 2024. After subtracting instrumental dead time, on-orbit calibration time, a giant solar flare between September 9  and September 13, 2017~\cite{solar}, and the South Atlantic anomaly passage time~\cite{orbitdata1}, the total exposure time was determined to be $2.17\times 10^{8}$ s, about 76\% of the total operational duration. 

The Geant4 toolkit (version 4.10.5)~\cite{geant4_1} was used to simulate the response of the DAMPE detector to different particles. For nickel nuclei with primary energies between 100 GeV and 10 TeV, the simulation implemented the FTFP\_BERT model, and for energy above 10 TeV, it implemented the EPOS-LHC model~\cite{geant4_2} which was linked to the GEANT4 with the cosmic ray Monte Carlo package~\cite{geant4_3,geant4_4}. Particles were generated from a hemispherical isotropic source surrounding the detector with a power-law energy spectrum (index $\gamma = -1$). The primary particle sample consisted of $\prescript{58}{}{\text{Ni}}$ and $\prescript{60}{}{\text{Ni}}$, mixed in a ratio of 5:2~\cite{isotope1,isotope2,isotope3}. Besides, there was also a simulation based on FLUKA 2011.2x~\cite{fluka} to evaluate systematic uncertainty from the hadronic interaction model.

\emph{Preselection}: To identify nickel candidate events from cosmic rays and reconstruct the kinetic energy spectrum, firstly the trajectories and charges of incident particles are required to be well reconstructed. Several selection criteria were applied to extract a high-purity sample of nickel candidates for kinetic energy reconstruction. In this analysis, deposited energy in BGO ($E_{\rm{dep}}$) was required to be more than 100 GeV, thereby avoiding the effects of the geomagnetic effect on low energy particles~\cite{rigidity}. Only events passing high energy trigger were analyzed~\cite{trigger}. For nickel with kinetic energy above 10 GeV/n, the trigger efficiency was nearly 100\%.

\emph{Track selection}: A method of track reconstruction using machine learning was developed to avoid efficiency loss for heavy nuclei caused by the saturation effect and common-mode noise compression in STK~\cite{mltrack}. This method maintains high efficiency even when the deposited energy in BGO exceeds 10 TeV. And the PSD charges along the track	 must exhibit the highest deposited energy among the adjacent four PSD strips in both the PSD-x and PSD-y layer. Additionally, particle trajectories must be incident on the top surface of the PSD detector and traverse the BGO calorimeter, through both its top and bottom surface. To suppress edge effects, the strip with the maximum deposited energy in each layer of the BGO calorimeter should not be the outmost one.

\begin{figure}[htbp]
\centering
\includegraphics[width=0.48\textwidth]{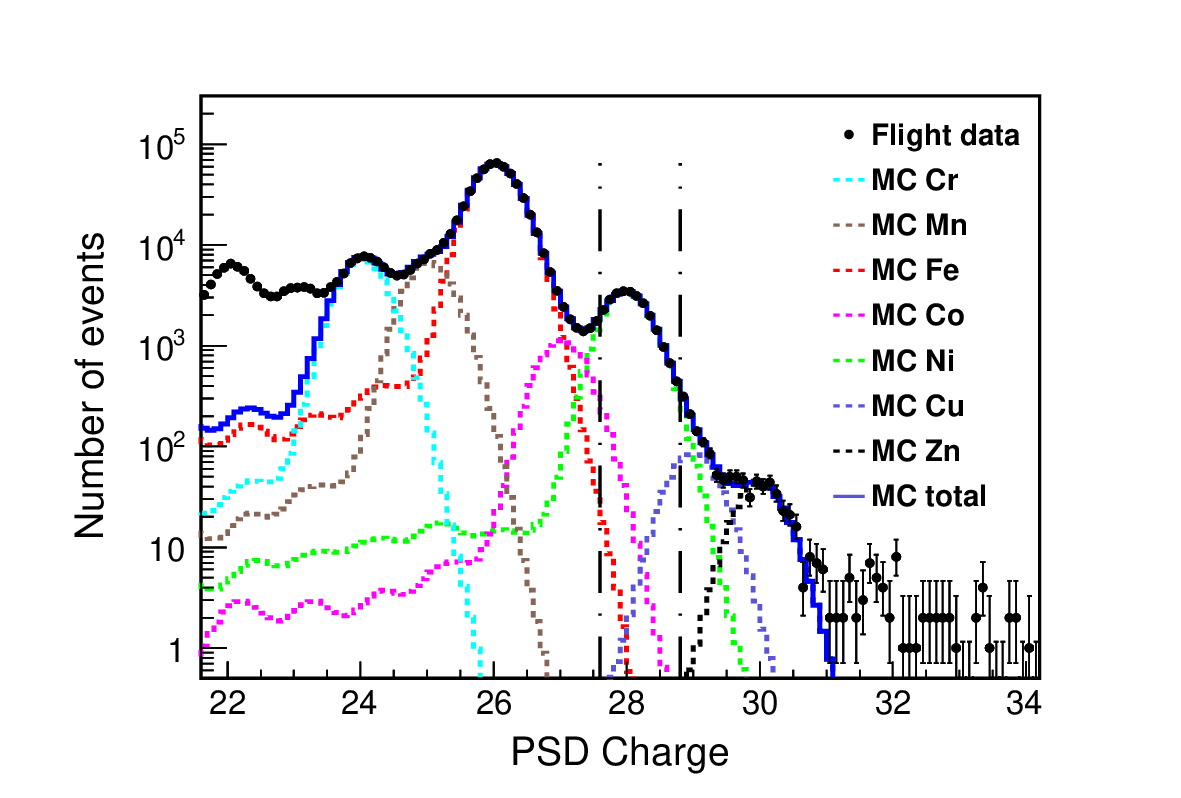}      
\includegraphics[width=0.48\textwidth]{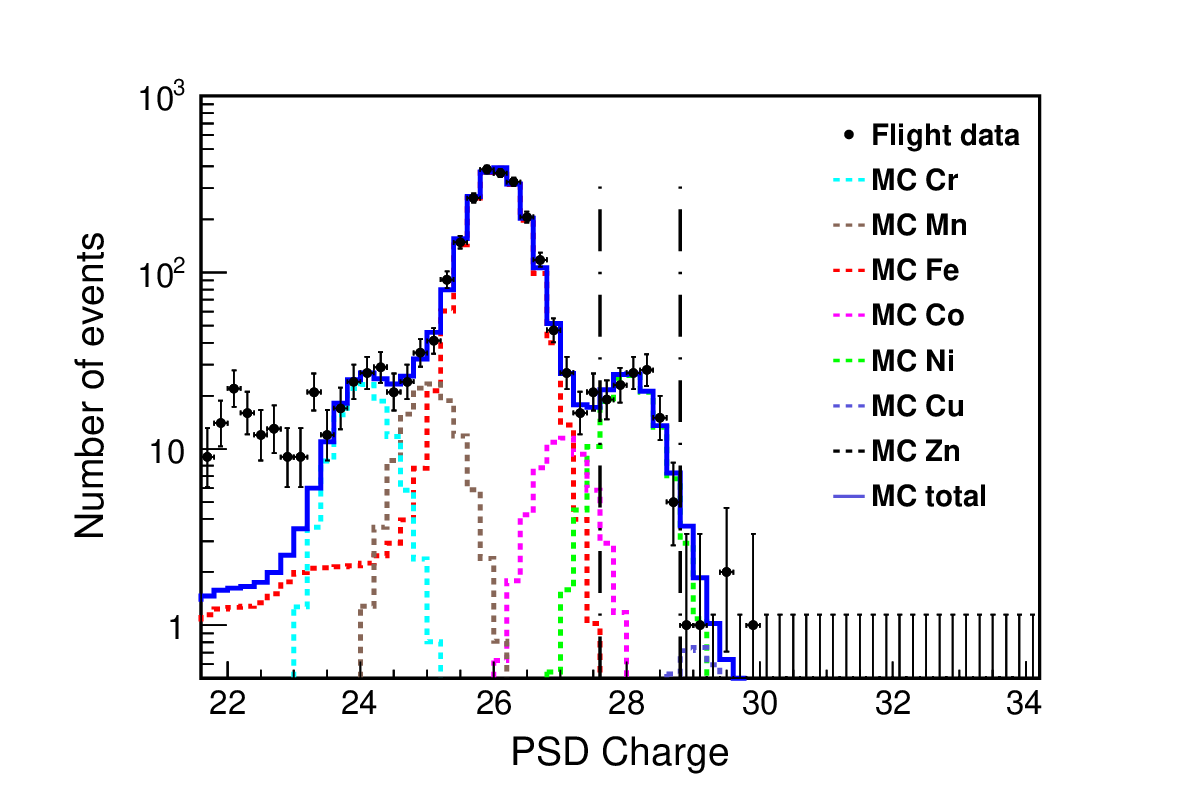}
\caption{\justifying The distribution of PSD charge of flight data (black points) within deposited energy bins of 100-158 GeV (top) and 4.64-10 TeV (bottom). 
The result of template fits is shown as the blue line, and the other nuclei as different colored dashed lines.}
\label{ChargeFit}
\end{figure}

\emph{Charge selection}: The PSD detector is mainly used for charge measurement. Charge reconstruction is based on ionization energy loss in PSD, scaling with the square of the particle charge according to the Bethe-Bloch formula. The charge value obtained has been corrected to eliminate position dependence, light attenuation, quenching effects and energy dependence~\cite{2019Tiekuang,orbitdata4}. Because the PSD detector is composed of four sublayers, two sublayers aligned in the yz view and two in the xz view, requiring particles to traverse at least three sublayers of the PSD detector is feasible to improve charge resolution. To guarantee the consistency of different PSD layers, the variance of these charge values in each PSD layer was required to be less than 1.2 as shown in Fig.S1 of the Supplemental Material (SM)~\cite{supple}\nocite{datapoint,nuisance}. A global PSD charge, detailed in the Supplemental Material~\cite{supple} with charge resolution of 0.37 e for nickel is used for charge identification. Fig.\ref{ChargeFit} shows the PSD charge distributions for deposited energy bins, 100-158 GeV and 4.64-10 TeV. The charge selection window for nickel was set from 27.6 to 28.8 (dot-dashed line in Fig.1), with approximately 80\% selection efficiency. A total of over 59 k candidate events were collected with deposited energy ranging from 100 GeV to 100 TeV.

\emph{Background subtraction}: To estimate the contamination from other nuclei, a template fit was performed based on the simulation. The MC charge spectra from chromium to zinc after smearing by convolved the simulated charge spectrum with a Gaussian function are also shown in Fig.\ref{ChargeFit} with different colored dashed lines. The dominant sources of contamination were iron, cobalt, and copper. Among these sources, cobalt was the highest, accounting for 2\% of the deposited energy at 100 GeV and 10\% at 100 TeV. The estimated contamination fractions from different nuclei are shown in Fig.S2 of the Supplemental Material~\cite{supple}. Contributions from other nuclei exhibited energy dependence, varying from 4\% of the deposited energy at 100 GeV to 15\% at 100 TeV. The background from nuclei heavier than zinc could be ignored because of low abundance.  

\emph{Effective acceptance}-The effective acceptance after selection in the \emph{i}th incident energy bin was calculated as 
\begin{eqnarray}
	A_{\text{eff},i} = A_{\text{gen}} \times \frac{N_{\text{pass},i}}{N_{\text{gen},i}}, \label{eq1}
\end{eqnarray} 

$A_{\text{gen}}$ is the geometrical factor of the MC events generator. In the \emph{i}th kinetic energy bin, $N_{\text{pass},i}$ is the number of events passing selection, and $N_{\text{gen},i}$ is the total number of events generated. The effective acceptance as the function of kinetic energy is approximately 0.07 $\text{m}^{2}\text{sr}$ above 10 GeV/n (see Fig.S3 in the Supplemental Material~\cite{supple}).

\emph{Energy measurement and unfolding procedure}-In this analysis, the total deposited energy in the BGO calorimeter was used for energy measurement. For a BGO crystal with energy deposited over 10 TeV, its readout electronics would saturate. To mitigate the impact of saturation effects on energy measurement, a correction algorithm based on MC simulation was performed~\cite{saturation}. The Birks' quenching effect in the BGO crystal was taken into account in the MC simulations~\cite{1951Birk,quenching}. This effect primarily affected the measurement of nickel spectra in the low energy region below 100 GeV/n.

Owing to the limited 1.6 nuclear interaction length of the BGO calorimeter, ion particles deposit approximately 30\% of their kinetic energy with a certain degree of broadening within the BGO calorimeter~\cite{BeamTest}. The relationship between the number of events in the \emph{i}th deposited energy bin $N(E_{\text{dep}}^i)$ and the number of events in the \emph{j}th incident kinetic energy bin $N(E_{\text{inc}}^j)$ is formalized as  
\begin{eqnarray}\label{eq2}
	N(E_{\text{dep}}^i) = \sum_{j=1}^{n} P(E_{\text{dep}}^i | E_{\text{inc}}^j ) N(E_{\text{inc}}^j),
\end{eqnarray}
where $P( E_{\text{dep}}^{i} | E_{\text{inc}}^{j})$ represents the energy response matrix, derived from MC simulations. A Bayesian unfolding method~\cite{unfold} was employed to convert the deposited energy spectrum back to the incident kinetic energy spectrum. 

\emph{Flux calculation}-The differential flux spectrum for nickel was calculated using the following formula:
\begin{eqnarray}
	\Phi_{i} = \Phi (E_{\text{inc}}^i,E_{\text{inc}}^i+\Delta E_i)= \frac{N_{\text{inc}}^i}{\Delta E_i T_{\text{exp}} A_{\text{eff},i}}, \label{eq3}
\end{eqnarray}
In \emph{i}th incident kinetic energy bin, $\Delta E_{i}$ is the width of the bin and $N_{\text{inc}}^i$ is the number of events after unfolding. $A_{\text{eff},i}$ is the effective acceptance and $T_{\text{exp}}$ is the exposure time. The fluxes are then expressed as kinetic energy per nucleon assuming the isotopic ratio of $\prescript{58}{}{\text{Ni}}$ : $\prescript{60}{}{\text{Ni}}$ = 5 : 2.

\emph{Uncertainty analysis}-The statistical uncertainty mainly originated from Poisson fluctuations of the number of observed events $N(E_{\text{dep}}^i)$. A toy-MC approach was applied in each deposited energy bin based on Poisson distribution. The statistical uncertainty was then evaluated by calculating the root-mean square of distribution in the corresponding kinetic energy bins.

The sources of systematic uncertainties in the nickel analysis are complex, including the selection efficiency, background subtraction, hadronic model, unfolding procedure and isotopic assumption. Comparing MC simulation with flight data, the difference was negligible for high energy trigger efficiency and within 0.5\% for track efficiency (see the Fig.S4 in the Supplemental Material~\cite{supple}). The uncertainty of charge selection was evaluated by adjusting the charge selection window from -0.1 to +0.2 corresponding to a variation of charge selection efficiency of 10\%. The flux in kinetic energy bins varied by ~1\% at 10 GeV/n and ~7\% at 1 TeV/n. The uncertainty of background subtraction was evaluated by varying the contamination level by 50\%. The differential flux of nickel would vary about 2\% at 10 GeV/n and 7\% at 1 TeV/n.

There are several selection criteria associated with charge reconstruction which should be considered into the source of systematic uncertainties. The systematic uncertainty associated with the requirement for matching the STK track with the PSD maximum energy strip was assessed by relaxing this condition and examining the variation of results. This caused a nickel flux variation less than 1\% at 10 GeV/n and 8\% at 1 TeV/n. A comparative study on the charge consistency efficiency was conducted showing that the difference between flight data and MC simulation was less than 4\% (see Fig.S1 in the Supplemental Material~\cite{supple}).           

The main source of systematic uncertainty below 200 GeV/n is the hadronic model. Below 75 GeV/n, the difference between the beam test data and simulation based on the FTFP\_BERT hadronic model in GEANT4 was attributed to the hadronic model uncertainty detailed in Fig.S5 and Fig.S6 of the Supplemental Material~\cite{supple}. Above 75 GeV/n, an independent analysis based on the DPMJET-III hadronic model in FLUKA was used to compare with the results based on FTFP\_BERT. This difference was about 6\% at 1 TeV/n. The systematic uncertainty from the unfolding procedure caused by the initial index and the MC sample in the unfolding matrix was estimated to be less than 1\%. Based on past isotopic observations, a variation of 15\% in the $\prescript{60}{}{\text{Ni}}$ abundance would bring a~0.8\%  difference into the result of nickel flux~\cite{isotope1,isotope2,isotope3}. Total uncertainties including statistical uncertainty and systematic uncertainties from different sources in different kinetic energy bins are shown in Fig.S7 of the Supplemental Material.

\begin{figure}[htbp]
\centering
\includegraphics[width=0.48\textwidth]{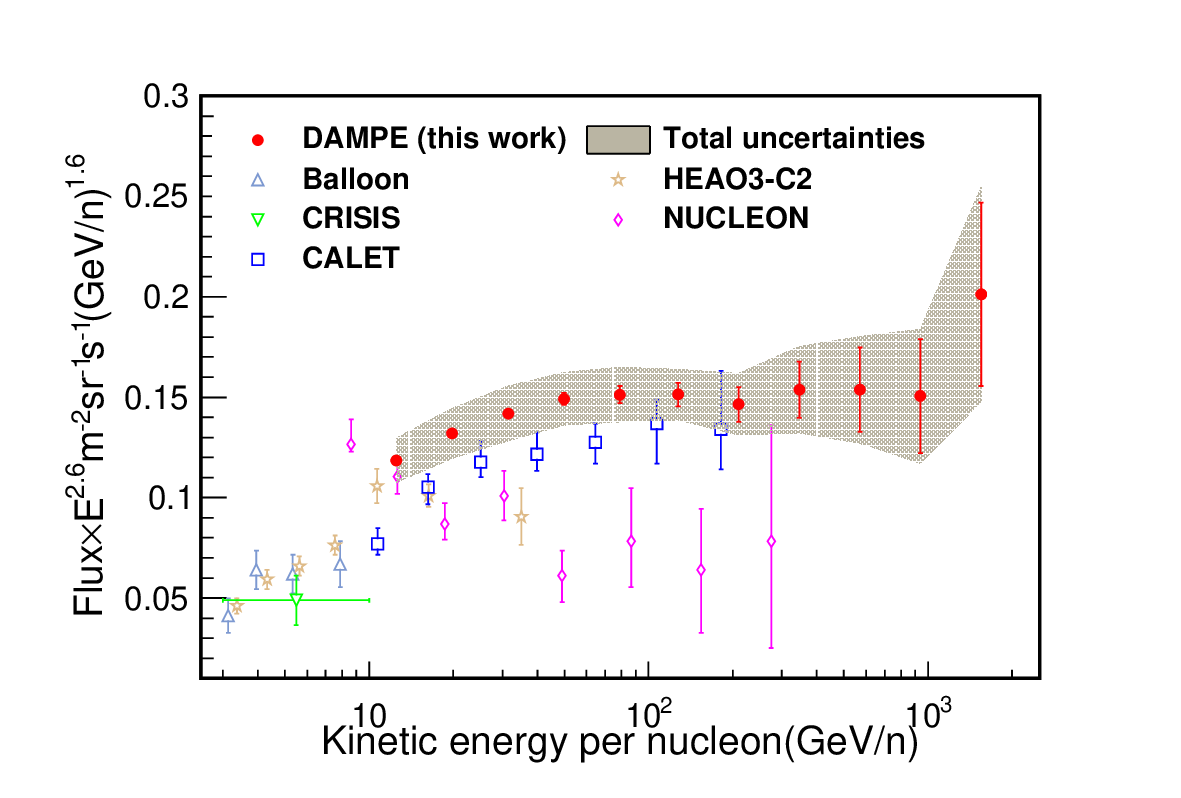}      
\includegraphics[width=0.48\textwidth]{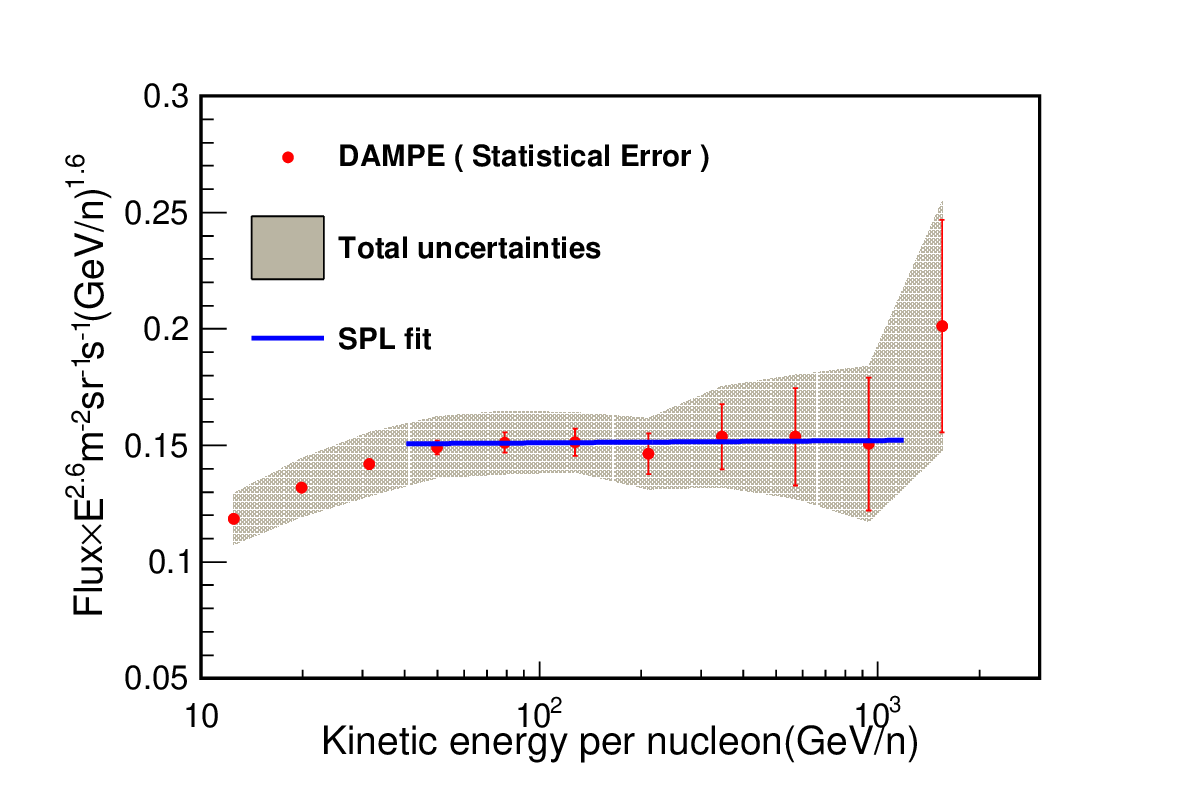}
\caption{\justifying Nickel spectrum weighted by $E^{2.6}$ measured by DAMPE (converted to kinetic energy per nucleon assuming the isotopic composition of A = 58.57). The nickel spectrum measured by DAMPE was compared with the previous results by CRISIS~\cite{1981Cris}, HEAO3-C2~\cite{1990HEAO3}, G. Minagawa~\cite{1981Balloon}, CALET~\cite{CALETNickel} and NUCLEON~\cite{NUCLEON}. The red bars represent the statistical uncertainty. The outer shaded bands represent the total uncertainties including statistical and systematic uncertainties from the analysis procedure and hadronic model. Bottom panel: the fit of the nickel spectrum  in the energy range [40 GeV/n, 1 TeV/n] with an SPL function (blue line). }\label{NickelSpectrum}
\end{figure}

\emph{Results}-Figure.~\ref{NickelSpectrum} shows the differential nickel spectrum in kinetic energy per nucleon from 10 GeV/n to 2 TeV/n. A softening feature is observed around 40 GeV/n which is similar to other nuclei like proton, helium or iron~\cite{2009Panov,2017Yoon,2011Adriani,2013Adriani,2015Aguilar,2020Aguilar,2015Aguilar_1,2017Aguilar,2021Aguilar,DAMPEProton,DAMPEHe,CALETProton,CALETIron}. A fit with a single power law function (SPL), 

\begin{eqnarray}\label{eq4}
	\Phi(E) = \Phi_{0} \left( \frac{E}{E_{0}} \right)^{\gamma}, \text{where}~ E_{0} = 1~\text{GeV/n},
\end{eqnarray}
was performed on the nickel flux with kinetic energy from 40 GeV/n to 1 TeV/n. The index $\gamma$ of the spectrum is $-2.60 \pm 0.03$ with a $\chi^2/\text{d.o.f} = 0.40/3$. Because of the limited statistics above 1 TeV/n, whether the nickel flux above TeV/n still follows a single power law remains to be verified.

\begin{figure}[htbp]
	\centering
	\includegraphics[width=0.48\textwidth] {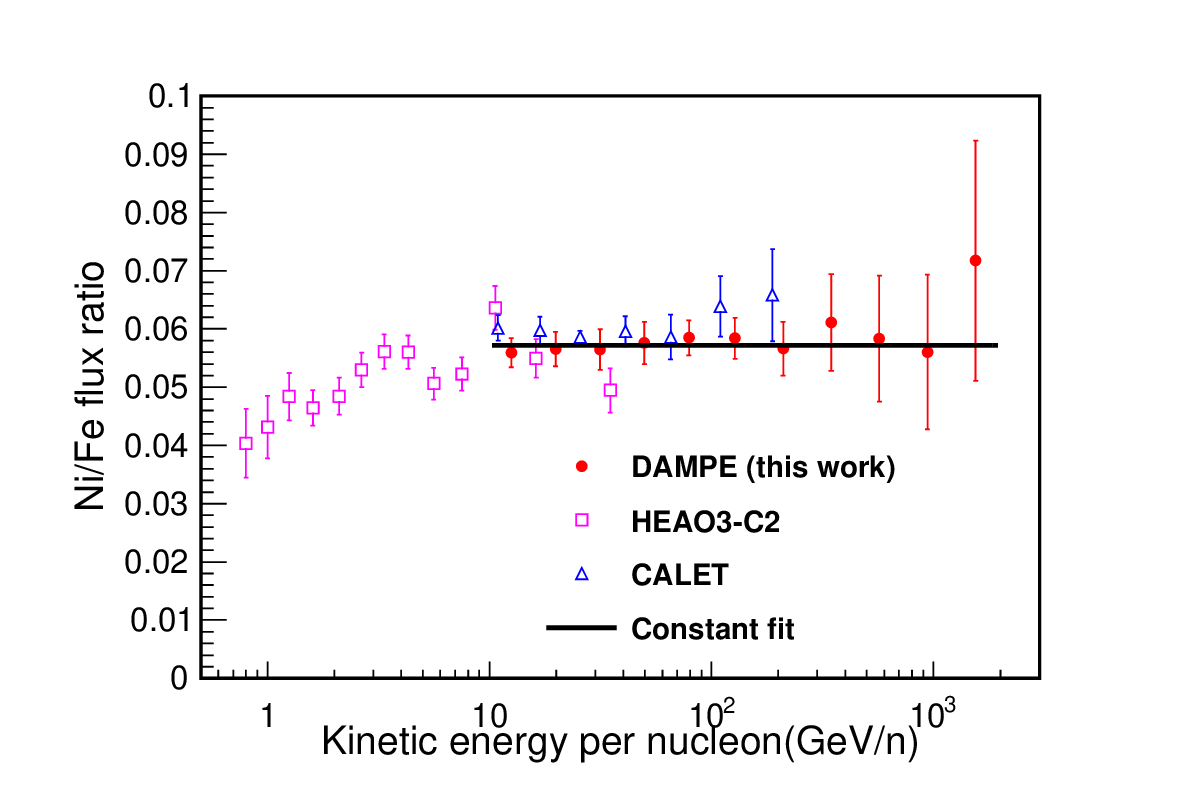}
     \justifying
	\caption{\justifying Nickel to iron flux ratio measured by DAMPE (red points) fit with a constant function (black line). The red bars represent the total uncertainty including statistical uncertainty and systematic uncertainties. The blue points show the measurements from CALET. The pink points shows the measurements from HEAO3-C2.}
	\label{ratio}
\end{figure}

The variation of the nickel to iron ratio with incident energy was also shown in Fig.~\ref{ratio}. A constant fit from 10 GeV/n to 2 TeV/n gives a constant value of $0.057 \pm 0.002$ with a $\chi^{2}/\text{d.o.f} = 2.12/7$.  
In this work, the nickel to iron ratio measured by DAMPE is structurally consistent with previous CALET measurements in the same energy range.

\emph{Conclusion}-In this Letter, the nickel spectrum from 10 GeV/n to 2 TeV/n is measured with nine years of data collected by DAMPE. There is a similar softening as lighter nuclei around 40 GeV/n in nickel flux. The nickel spectrum measured by DAMPE is well consistent with a SPL fit above 40 GeV/n which gives a spectra index of $\gamma = -2.60 \pm 0.03$. In the energy region from 10 GeV/n to 2 TeV/n, the nickel to iron ratio can be approximately described by a constant $0.057 \pm 0.002$. In theory, iron and nickel are mainly synthesized thermonuclear explosions of supernovae, as well as in Si burning during the explosive burning of core-collapse supernovae~\cite{motivation1,motivation2,motivation3,motivation4}. The constant Ni/Fe abundance ratio in cosmic rays further indicates the similar origin, acceleration, and propagation mechanism for both elements. A hardening structure was observed for iron at kinetic energy 512.5 GeV/n~\cite{softening}. For nickel at kinetic energy above hundreds of GeV/n, current statistics are not enough to draw a definitive conclusion. Our measurements represent the first observation of the nickel energy spectrum extending to TeV/n with a higher precision than most of the existing experimental results. The accumulating DAMPE dataset would facilitate more precision measurements of nickel extending to higher energy.

\emph{Acknowledgment}-The DAMPE mission was funded by the strategic priority science and technology projects in space science of the Chinese Academy of Sciences (CAS). In China, the data analysis is supported by the National Key Research and Development Program of China (No.  2022YFF0503303) and the National Natural Science Foundation of China (No. 12275266,No. 12588101,No. 12220101003,No. 12022503, and No. 12503107),  the  CAS  Project  for  Young  Scientists in  Basic  Research  (No. YSBR-061  and No. YSBR-092),  the  Strategic  Priority  Program  on Space Science of Chinese Academy of Sciences (No.  E02212A02S), the Youth Innovation Promotion Association of CAS, the Young Elite Scientists Sponsorship Program by CAST (No. YESS20220197), and the Program for Innovative Talents and Entrepreneur in Jiangsu. Y.Z. F is thankful for the support of New Cornerstone Science Foundation through XPLORERPRIZE.  In  Europe,  the  activities  and  data  analysis  are  supported  by  the  Swiss  National Science Foundation (SNSF), Switzerland, the National Institute for Nuclear Physics (INFN), Italy, and the European Research Council (ERC) under the European Union’s Horizon 2020 research and innovation program (No.  851103), and the Swiss State Secretariat for Education, Research and Innovation (SERI).

\emph{Data availability}-The data that support the findings of this article are openly available\cite{dataset}.

\bibliographystyle{apsrev4-2}
\bibliography{mybibfile}

\end{document}